\documentclass[preprint,prd,showpacs,superscriptaddress,preprintnumbers]{revtex4}

\usepackage{amsmath}
\usepackage{amssymb}
\usepackage{enumerate}
\usepackage{bm}
\usepackage{dcolumn}
\usepackage{graphicx}
\usepackage{enumerate}
\usepackage{bibentry}

\usepackage{subfigure}

\newenvironment{resulttable}{%
\begin{tabular}{lc@{\hskip-3em}d@{\hskip2em}c@{\hskip1em}c}
\makebox[7em][l]{Integral} & $n_F$ & \makebox[-5em]{Value (Error)}   & Sampling per & No. of   \\ [-.5ex]  
                           &       & \makebox[-5em]{including $n_F$} & iteration    & iterations   \\ 
\hline
}{%
\end{tabular}
}

\newenvironment{renomtable}{%
\begin{tabular}{l@{\hskip-3em}d@{\hskip4em}l@{\hskip-3em}d}
Integral & \makebox[-4em]{Value(Error)} & 
Integral & \makebox[-4em]{Value(Error)}   \\
\hline
}{%
\end{tabular}
}

\begin{document}

\date{\today}

\preprint{NTLP 2008-02}

\title{%
Tenth-Order Lepton Anomalous Magnetic Moment ---
Second-Order Vertex Containing Two Vacuum Polarization Subdiagrams,
One Within the Other
}


\author{Tatsumi Aoyama}
\affiliation{Institute of Particle and Nuclear Studies,
High Energy Accelerator Research Organization (KEK),
Tsukuba, Ibaraki 305-0801, Japan}

\author{Masashi Hayakawa}
\affiliation{Department of Physics, Nagoya University, Nagoya, Japan 464-8602 }

\author{Toichiro Kinoshita}
\affiliation{Laboratory for Elementary-Particle Physics, Cornell University, Ithaca, New York, 14853, U.S.A }

\author{Makiko Nio}
\email{nio@riken.jp}
\affiliation{Theoretical Physics Laboratory, Nishina Center, RIKEN, Wako, Japan 351-0198 }

\begin{abstract}
This paper reports 
the tenth-order QED contribution to the $g\!-\!2$ of electron and muon from
two gauge-invariant sets, Set I($g$) and Set I($h$). 
In the case of electron $g\!-\!2$
Set I($g$) consists of
nine Feynman diagrams which have a fourth-order
vacuum-polarization loop containing another fourth-order
vacuum-polarization loop.
Set I($h$) 
consists of 30 Feynman diagrams which have a proper sixth-order
vacuum-polarization loop containing a second-order
vacuum-polarization loop.
The results of numerical integration,
including mass-dependent terms containing one closed loop
of muon, are 0.028~597~(4) $(\alpha/\pi)^5$ for Set I($g$)
and 0.001~685~(13) $(\alpha/\pi)^5$  for Set I($h$), respectively.
We also report the contributions of Set I($g$) and Set I($h$)
to the muon anomaly.
Diagrams included are those containing electron, muon, and tau-lepton loops.
Their sums are $2.640~9~(4) (\alpha/\pi)^5$  and $-0.564~8~(11) (\alpha/\pi)^5$, respectively.
The sum of contributions of Sets I($g$) and I($h$) containing only electron loops
are in fair agreement with the
recently obtained asymptotic analytic results.

\end{abstract}

%


\maketitle

\section{Introduction}
\label{sec:intro}

The anomalous magnetic moment $g\!-\!2$ of the electron has played the central role
in testing the validity of QED.
To match the precision of the recent measurement of the electron $g\!-\!2$ \cite{odom},
the theory must include the radiative correction of up to the eighth-order \cite{kn1,ahkn1},
the hadronic contribution \cite{davier,HMNT, krause,melnikov},
and the electroweak contribution
\cite{czarnecki,knecht,marciano}
within the context of the Standard Model.
As a matter of fact, the largest theoretical uncertainty now comes from
the not-yet-calculated tenth-order term.
Thus, for a more stringent test of QED, it is necessary to know,
not a crude estimate made in Ref. \cite{mohr},
but an actual value of the tenth-order term.
To meet this challenge
we launched several years ago a systematic program to evaluate the complete tenth-order term
\cite{kn2,aoyama1,aoyama2}.

The tenth-order QED contribution to the 
anomalous magnetic moment of an electron can be written as
\begin{equation}
	a_e^{(10)} 
	= \left ( \frac{\alpha}{\pi} \right ) ^5 \left [A_1^{(10)}
	+ A_2^{(10)} (m_e/m_\mu) 
	+ A_2^{(10)} (m_e/m_\tau) 
	+ A_3^{(10)} (m_e/m_\mu, m_e/m_\tau) \right ] .
\label{eq:ae10th}
\end{equation}
The contribution to the mass-independent term $A_1^{(10)}$ may be
classified into six gauge-invariant sets, further divided into
32 gauge-invariant subsets depending on the nature of closed
lepton loop subdiagrams.
Thus far, results of numerical evaluation of 18 gauge-invariant subsets, 
which consist of 964 vertex diagrams, 
have been published \cite{kn2,watanabe}.
Some of 18 subsets were also analytically calculated \cite{Laporta} 
and show good agreement with the numerical results.
Several more gauge-invariant sets have been evaluated
and are being prepared for publication.

\begin{figure}
\includegraphics[width=12cm]{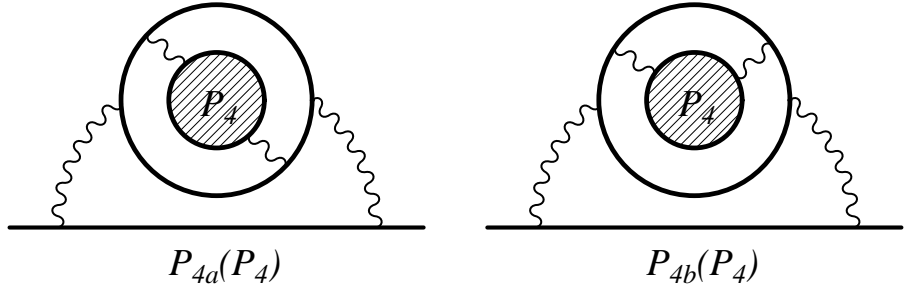}
\caption{Typical diagrams of Set I($g$) which contain 
eighth-order vacuum-polarization subdiagrams $\Pi^{(4,4)} (q^2)$.
Nine diagrams belong to this set.}
\label{fig:vp4}
\end{figure}

\begin{figure}
\includegraphics[width=15cm]{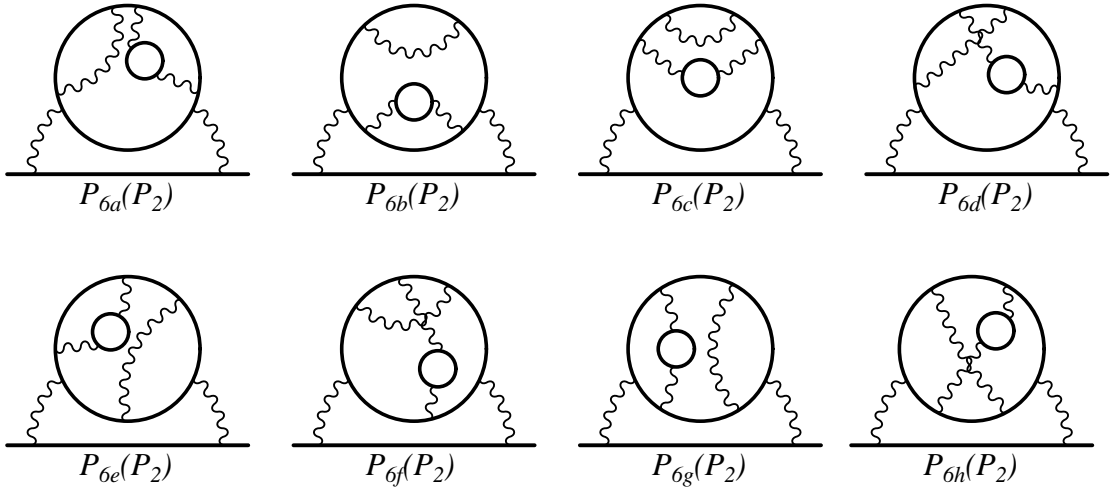}
\caption{Typical diagrams of Set I($h$) which contain 
eighth-order vacuum-polarization diagrams $\Pi^{(6,2)} (q^2)$.
Thirty diagrams belong to this set.}
\label{fig:vp6}
\end{figure}

In this paper we report the contribution to $A_1^{(10)}$ from
two gauge-invariant subsets, called Set I($g$) and Set I($h$).
Set I($g$) consists of diagrams which contain 4th-order
vacuum-polarization diagrams $\Pi^{(4)}$ 
whose internal photon line  has an insertion of another 4th-order
vacuum-polarization loop. 
This is denoted as $\Pi^{(4,4)}$.  (See Fig. \ref{fig:vp4}.)
Set I($h$) consists of diagrams which contain proper 6th-order
vacuum-polarization diagrams $\Pi^{(6)}$ 
in whose internal photon lines a  second-order
vacuum-polarization loop is inserted.
This is denoted as $\Pi^{(6,2)}$.
(See Fig. \ref{fig:vp6}.)

The evaluation of the contributions of Set I($g$) and Set I($h$)
to $a_e^{(10)}$ is straightforward if the spectral functions
of $\Pi^{(4,4)}$ and $\Pi^{(6,2)}$ are known,
since contributions of these vacuum-polarization loops
 can be regarded as superpositions of the second-order
$g\!-\!2$ in which the virtual photon is replaced by massive vector bosons
whose distribution is weighted by the spectral function.
Unfortunately this approach is not fully applicable to our problem since
exact spectral functions are known only for 
$\Pi^{(2)}$, $\Pi^{(4)}$ \cite{kallen}, and $\Pi^{(4,2)}$  \cite{hoang}.
Instead of pursuing the photon spectral function, we therefore follow
an alternative approach \cite{QEDbook} in which 
eighth-order vacuum-polarization functions
$\Pi^{(4,4)}$ and $\Pi^{(6,2)}$ are constructed from 
Feynman-parametric integrals of $\Pi^{(4)}$ and $\Pi^{(6)}$, 
which are then inserted in the virtual photon line of the Feynman-parametric integral of
the second-order anomalous magnetic moment $M^{(2)}$.
This procedure leads to the formula \cite{QEDbook}
\begin{equation}
 M_{2,P}= - \int_0^1 dy (1-y) \Pi \left (q^2 \right )|_{q^2= - \frac{y^2}{1-y} }, 
\label{m2p}
\end{equation}
where $\Pi(q^2)$ represents the vacuum-polarization function inserted into the photon propagator
with the momentum $q$.

There are two ways to construct $\Pi^{(4,4)}$.
One is to treat the entire eighth-order function by
direct Feynman parametrization.
Another is to insert the exact fourth-order spectral function
of $\Pi^{(4)}$ \cite{kallen} in the Feynman-parametric
integral of $\Pi^{(4)}$.
We choose the second approach in this paper.

For Set I($h$) an exact spectral function is not yet known.
Thus we follow the alternative approach \cite{QEDbook} 
which utilizes the parametric representation of the vacuum-polarization function
$\Pi^{(6)}$ itself.
Since the parametric integral of $\Pi^{(6)}$ is known \cite{QEDbook},
it is easy to obtain $\Pi^{(6,2)}$
by insertion of the second-order 
vacuum-polarization function or its spectral function in $\Pi^{(6)}$.
For simplicity we choose the latter approach.

Renormalization of the vacuum-polarization function is carried out
in two steps.
The first step is to remove the
UV divergences coming from subdiagrams of the vacuum-polarization function
by the K-operation \cite{QEDbook},
supplemented by the R-subtraction \cite{aoyama2} for some diagrams.
Then a UV-finite correction, called residual renormalization, is applied
to achieve the standard on-the-mass-shell renormalization.


Diagrams discussed in this paper consist of an open lepton line ($l_1$),
a closed lepton line ($l_2$), and another closed lepton line ($l_3$)
inserted in an internal photon line of ($l_2$), where ($l_i$) represents electron ($e$),
muon ($m$), or tau lepton ($t$).
Thus each diagram is characterized by a superscript $(l_1l_2l_3)$.
Residual renormalization constants are also denoted by superscripts
such as $(l_2l_3)$, where $l_2$ is an external lepton line and $l_3$ refers to the inner loop.

The evaluation of Set I($g$) 
is described in Sec. \ref{sec:setIg}.
The evaluation of Set I($h$) 
is described in Sec. \ref{sec:setIh}.
Sec. \ref{sec:summary} is devoted to the summary and discussion
of our results.
For simplicity the factor $(\alpha/\pi)^5$ is omitted in Secs. \ref{sec:setIg}
and \ref{sec:setIh}.


\section{Set {\rm I}($g$)}
\label{sec:setIg}

In this section we consider Set I($g$).  
The renormalized contribution of Set I($g$) to the magnetic moment
of the lepton $l_1$ is given by  the general formula
\cite{QEDbook}
\begin{equation}
a_{l_1}^{(10)} [{\rm I}(g)^{(l_1l_2l_3)}] = \Delta M_{2,P4a(P4)}^{(l_1l_2l_3)} + 2\Delta M_{2,P4b(P4)}^{(l_1l_2l_3)}  
       - 2 \Delta B_{2,P4}^{(l_2l_3)}  M_{2,P2}^{(l_1l_2)},
\label{set1g}
\end{equation}
where the first two terms are finite integrals obtained by the K-operation and 
the last term is the residual renormalization term.  


\subsection{Set ${\rm I}(g)^{(eee)}$}
\label{subsec:setIgeee}

Let us first consider
the case where $l_1, l_2, l_3$ are all
electron lines, namely $(l_1 l_2 l_3)=(eee)$:
\begin{equation}
a_e^{(10)}[{\rm I}(g)^{(eee)}] = \Delta M_{2,P4a(P4)}^{(eee)}
         + 2\Delta M_{2,P4b(P4)}^{(eee)}
       - 2 \Delta B_{2,P4}^{(ee)} M_{2,P2}^{(ee)}.
\label{set1geee}
\end{equation}
This gives a mass-independent contribution to the electron $g\!-\!2$.
The numerical values of
$\Delta M_{2,P4a(P4)}^{(eee)}$ and $\Delta M_{2,P4b(P4)}^{(eee)}$ 
obtained by the Monte Carlo integration routine VEGAS \cite{vegas}
are listed in
Table \ref{tablege} and those of $\Delta B_{2,P4}^{(ee)}$ and $M_{2,P2}^{(ee)}$ are listed in Table \ref{tablexxaux}.
Inserting these values 
in Eq. (\ref{set1geee}) we obtain 
\begin{equation}
a_e^{(10)} [{\rm I}(g)^{(eee)}] = 0.028~569~(6).
\label{eq:Igeee}
\end{equation}


\renewcommand{\arraystretch}{0.80}
\begin{table}
\begin{ruledtabular}
\caption{ Contributions of diagrams
of Set I($g$) of Fig. 1 to the electron $g\!-\!2$ with $(eee)$, $(eem)$, and $(eme)$
and muon $g\!-\!2$ with $(mee)$, $(mem)$, and $(mme)$.
$n_F$ is the number of Feynman diagrams represented by the integral.
All integrals are evaluated by VEGAS \cite{vegas} in double precision.
  \\
\label{tablege}
}
\begin{resulttable}
  \\
$\Delta M_{2,P4a(P4)}^{(eee)}$&3&0.028~534~(5)&  $1 \times 10^7$ & 50   \\
$\Delta M_{2,P4b(P4)}^{(eee)}$&6&0.005~798~(2)&  $1 \times 10^7$ & 50   \\
  \\
$\Delta M_{2,P4a(P4)}^{(eem)}$&3&0.171~9~(2)\times 10^{-4} &$1 \times 10^6$ &20   \\
$\Delta M_{2,P4b(P4)}^{(eem)}$&6&0.002~1~(0)\times 10^{-4} &$1 \times 10^6$ &20   \\
  \\
$\Delta M_{2,P4a(P4)}^{(eme)}$&3&0.052~2~(1)\times 10^{-4}&$1 \times 10^6$ &20   \\
$\Delta M_{2,P4b(P4)}^{(eme)}$&6&0.044~6~(1)\times 10^{-4}&$1 \times 10^6$ &20   \\
  \\
$\Delta  M_{2,P4a(P4)}^{(mee)} $ &3& 0.533~54~( 16)  & $ 1\times 10^7, 1\times 10^8$ &
150, 50   \\
$\Delta  M_{2,P4b(P4)}^{(mee)} $ &6& 2.219~96~( 17)  & $ 1\times 10^7, 1\times 10^8$ &
150, 50   \\
  \\
$\Delta  M_{2,P4a(P4)}^{(mem)} $ &3& 0.059~619~( 60) & $ 1\times 10^7$ &     20   \\
$\Delta  M_{2,P4b(P4)}^{(mem)} $ &6& 0.002~157~( 12) & $ 1\times 10^7$ &     20   \\
  \\
$\Delta  M_{2,P4a(P4)}^{(mme)} $ &3& 0.139~86~( 19)  & $ 1\times 10^7$ &     20   \\
$\Delta  M_{2,P4b(P4)}^{(mme)} $ &6& 0.131~30~( 23)  & $ 1\times 10^7$ &     20   \\
  \\
\end{resulttable}
\end{ruledtabular}
\end{table}
\renewcommand{\arraystretch}{1}

\renewcommand{\arraystretch}{0.80}
\begin{table}
\begin{ruledtabular}
\caption{ Auxiliary integrals for  Sets I($g$) and I($h$).
Some integrals are known exactly. 
Other integrals are obtained by the integration routine VEGAS.
The superscript $P4$ stands for the sum of the parametric constructions 
of the fourth-order vacuum-polarization functions $P4a$ and $P4b$.
For the case of $\Delta B_{2,P4}$, the fourth-order vacuum polarization is inserted
into $\Delta B_2$ by using the K\"{a}ll\'{e}n-Sabri spectral function.  
The relation between $P4$ here and the sum of $P4a$ and  $P4b$ is given by 
$\Delta B_{2,P4}=\Delta B_{2,P4a}+\Delta B_{2,P4b} - \Delta B_2 \Delta B_{2,P2}$. 
\label{tablexxaux}
} 
\begin{renomtable}
  \\
$M_{2,P2}^{(ee)}$                   &  0.015~687~421 ... &
$\Delta M_{2,P4}^{(ee)}$            &  0.076~401~785 ...    \\
$M_{2,P2}^{(em)}$                   &  0.005~197~(1)\times 10^{-4} & 
$\Delta M_{2,P4}^{(em)}$            &  0.027~526~(1) \times 10^{-4}     \\
$M_{2,P2}^{(me)}$                   &  1.094~259~6~(0)  &
$\Delta M_{2,P4}^{(me)}$            &  3.134~97~(14)    \\
  \\
$\Delta B_{2}$                      &  0.75                      & 
                                    &                              \\     
$\Delta B_{2,P2}^{(ee)}$            &  0.063~399~266 ...          &
$\Delta B_{2,P4}^{(ee)}$            &  0.183~666~5~(18)             \\ 
$\Delta B_{2,P2}^{(em)}$            &  0.094~054~(1)\times 10^{-4}  & 
$\Delta B_{2,P4}^{(em)}$            &  0.338~738~(12) \times 10^{-4}   \\ 
$\Delta B_{2,P2}^{(me)}$            &  1.885~766~(77)               &
$\Delta B_{2,P4}^{(me)}$            &  2.438~91~(23)                 \\
  \\
$\Delta M_{2,P4(P2)}^{(eee)}$       &  0.013~120~(1)  &
$\Delta M_{2,P4(P2)}^{(eem)}$       &  0.047~678~(13)\times 10^{-4}   \\
$\Delta M_{2,P4(P2)}^{(eme)}$       &  0.078~454~(8)\times 10^{-4}  &
$\Delta M_{2,P4(P2)}^{(mee)}$       &  1.579~51~(4)    \\
  \\
$\Delta B_{4,P2}^{(ee)}$            & -0.314~320~(10)    &
$\Delta L_{4,P2}^{(ee)}$            &  0.200~092~(14)      \\ 
$\Delta B_{4,P2}^{(em)}$            & -0.915~7~(47)\times 10^{-4}    &
$\Delta L_{4,P2}^{(em)}$            &  0.116~1~(64)  \times 10^{-4}    \\ 
$\Delta B_{4,P2}^{(me)}$            & -3.420~4~(72)    &
$\Delta L_{4,P2}^{(me)}$            &  3.121~3~(63)      \\ 
  \\
\end{renomtable}
\end{ruledtabular}
\end{table}
\renewcommand{\arraystretch}{1}

\subsection{Set ${\rm I}(g)^{(eem)}$}
\label{subsec:setIgeem}

The contribution of Set I($g$) to $a_e$,
in which the inner vacuum-polarization consists of muon loop,
is given by
\begin{equation}
a_e^{(10)} [{\rm I}(g)^{(eem)}] = \Delta M_{2,P4a(P4)}^{(eem)}
         + 2\Delta M_{2,P4b(P4)}^{(eem)}
       - 2 \Delta B_{2,P4}^{(em)} M_{2,P2}^{(ee)},
\label{set1geem}
\end{equation}
where  the numerical values of
$\Delta M_{2,P4a(P4)}^{(eem)}$ and $\Delta M_{2,P4b(P4)}^{(eem)}$ 
are listed in
Table \ref{tablege} and those of $\Delta B_{2,P4}^{(em)}$ 
and $M_{2,P2}^{(ee)}$ are listed in Table \ref{tablexxaux}.  
$m_\mu/m_e = 206.768~282~3~(52)$ is from Ref. \cite{mohr}.
Inserting these values 
in Eq. (\ref{set1geem}) we obtain 
\begin{equation}
a_e^{(10)} [{\rm I}(g)^{(eem)}] = 0.163~4~(2)\times 10^{-4}.
\label{eq:Igeem}
\end{equation}


\subsection{Set ${\rm I}(g)^{(eme)}$}
\label{subsec:setIgeme}

The contribution of Set I($g$) to $a_e$,
in which the outer vacuum-polarization consists of muon loop,
is given by
\begin{equation}
a_e^{(10)} [{\rm I}(g)^{(eme)}] = \Delta M_{2,P4a(P4)}^{(eme)}
         + 2\Delta M_{2,P4b(P4)}^{(eme)}
       - 2 \Delta B_{2,P4}^{(me)} M_{2,P2}^{(em)},
\label{set1geme}
\end{equation}
where  the numerical values of
$\Delta M_{2,P4a(P4)}^{(eme)}$ and $\Delta M_{2,P4b(P4)}^{(eme)}$ 
are listed in
Table \ref{tablege} and those of $\Delta B_{2,P4}^{(me)}$ 
and $M_{2,P2}^{(em)}$ 
are listed in Table \ref{tablexxaux}.  
Inserting these values 
in Eq. (\ref{set1geme}) we obtain 
\begin{equation}
a_e^{(10)} [{\rm I}(g)^{(eme)}] = 0.071~5~(1)\times 10^{-4}.
\label{eq:Igeme}
\end{equation}

Terms for $(emm)$, $(eet)$, etc., are even smaller
than those for $(eem)$ and $(eme)$.
Of course, they are easy to evaluate, if needed.
However, they are of no physical significance at present.



\subsection{Contribution to muon $g\!-\!2$ from  Set ${\rm I}(g)$ }
\label{subsec:setIgmee}

The leading contribution of Set I($g$) to the muon anomaly $a_\mu$
comes from the diagrams
in which both vacuum-polarization loops consist of electrons, namely,
\begin{equation}
a_\mu^{(10)} [{\rm I}(g)^{(mee)}] = \Delta M_{2,P4a(P4)}^{(mee)}
         + 2\Delta M_{2,P4b(P4)}^{(mee)}
       - 2 \Delta B_{2,P4}^{(ee)} M_{2,P2}^{(me)},
\label{set1gmee}
\end{equation}
where  the numerical values of
$\Delta M_{2,P4a(P4)}^{(mee)}$ and $\Delta M_{2,P4b(P4)}^{(mee)}$ 
are listed in
Table \ref{tablege}
and those of $\Delta B_{2,P4}^{(ee)}$ and $M_{2,P2}^{(me)}$ 
are listed in Table \ref{tablexxaux}.
Inserting these values 
in Eq. (\ref{set1gmee}) we obtain 
\begin{equation}
a_\mu^{(10)} [{\rm I}(g)^{(mee)}] = 2.351~53~(23).
\label{eq:amugee}
\end{equation}

Contributions of terms of type $(mme)$  
and $(mem)$ are
\begin{equation}
a_\mu^{(10)} [{\rm I}(g)^{(mme)}] = 0.194~64~(29),
\label{eq:amugme}
\end{equation}
and
\begin{equation}
a_\mu^{(10)} [{\rm I}(g)^{(mem)}] = 0.061~702~(61),
\label{eq:amugem}
\end{equation}
respectively.

Terms involving tau-lepton loop are three orders of magnitude 
smaller than (\ref{eq:amugee}), but greater than the uncertainty of (\ref{eq:amugee}).
Contributions of terms of type $(met)$, $(mte)$, $(mmt)$, $(mtm)$, and $(mtt)$ are
\begin{align}
& a_\mu^{(10)} [{\rm I}(g)^{(met)}] = 0.001~236~(2),  
\label{eq:amuget}
 \\
& a_\mu^{(10)} [{\rm I}(g)^{(mte)}] = 0.001~485~(4),  
\label{eq:amugte} \\
& a_\mu^{(10)} [{\rm I}(g)^{(mmt)}] = 0.000~893~(2),  
\label{eq:amugmt} \\
& a_\mu^{(10)} [{\rm I}(g)^{(mtm)}] = 0.000~663~(1),  
\label{eq:amugtm} \\
& a_\mu^{(10)} [{\rm I}(g)^{(mtt)}] = 0.000~141~(1),  
\label{eq:amugtt}
\end{align}
respectively.


\section{Set {\rm I}($h$)}
\label{sec:setIh}

For Set I($h$) the formula for the
 renormalized quantity, including residual renormalization,
takes different forms depending on whether one follows the original K-operation
\cite{QEDbook},
which subtracts only the UV-divergent part of mass-renormalization constant,
or the recently developed R-subtraction method \cite{aoyama2},
which subtracts the entire mass-renormalization term.
We follow here the latter approach which affects the definition of
$\Delta M_{2,P6i(P2)}^{(l_1l_2l_3)}$ for $i=c, d$.
Of course, these two methods 
are equivalent and give the same results.

\subsection{Set ${\rm I}(h)^{(eee)}$}
\label{subsec:setIheee}

The renormalized contribution of Set I($h$) with $(eee)$
to $a_e$ is given by
\begin{eqnarray}
a_e^{(10)} [{\rm I}(h)^{(eee)}] &=& \sum_{i=a}^h \Delta M_{2,P6i(P2)}^{(eee)}
\nonumber   \\
 &-&4 \Delta B_2 \Delta M_{2,P4(P2)}^{(eee)}-4 \Delta B_{2,P2}^{(ee)} \Delta M_{2,P4}^{(ee)} 
\nonumber   \\
 &+& 10 \Delta B_2 \Delta B_{2,P2}^{(ee)} M_{2,P2}^{(ee)}  
- 2 (\Delta L_{4,P2}^{(ee)} + \Delta B_{4,P2}^{(ee)}) M_{2,P2}^{(ee)},
\label{set1h}
\end{eqnarray}
where $\Delta M_{2,P4(P2)}^{(eee)}$, $\Delta L_{4,P2}^{(ee)}$,
and $\Delta B_{4,P2}^{(ee)}$
are obtained from the lower-order relations \cite{QEDbook}
\begin{eqnarray}
\Delta M_{2,P4} &=& \Delta M_{2,P4a} + 2\Delta M_{2,P4b}, 
\nonumber   \\
\Delta L_4 &=& \Delta L_{4x} + 2\Delta L_{4c} + \Delta L_{4l} + 2 \Delta L_{4s}. 
\nonumber   \\
\Delta B_4 &=& \Delta B_{4a} + \Delta B_{4b}, 
\end{eqnarray}
by insertion of a vacuum-polarization loop $P2$
in each of two photon lines of $P_{4a}$, 
$P_{4b}$, $\Delta L_4$, and $\Delta B_4$.
Substituting numerical values listed in Tables \ref{tablexxaux} and \ref{tablehe}
in Eq. (\ref{set1h}) we obtain 
\begin{equation}
a_e^{(10)} [{\rm I}(h)^{(eee)}] = 0.001~696~(13).
\label{eq:Iheee}
\end{equation}


\renewcommand{\arraystretch}{0.80}
\begin{table}
\begin{ruledtabular}
\caption{ Contributions of diagrams
of Set I($h$) of Fig.2 with $(eee)$, $(eem)$, and $(eme)$.
$n_F$ is the number of Feynman diagrams represented by the integral.
All integrals are evaluated in double precision.
  \\
\label{tablehe}
}
\begin{resulttable}
  \\
$\Delta M_{2,P6a(P2)}^{(eee)}$&4&0.003~269~(2)&$1 \times 10^8$ &150   \\
$\Delta M_{2,P6b(P2)}^{(eee)}$&2&0.002~367~(2)&$1 \times 10^8$ &150   \\
$\Delta M_{2,P6c(P2)}^{(eee)}$&4&-0.003~602~(2)&$1 \times 10^8$ &150   \\
$\Delta M_{2,P6d(P2)}^{(eee)}$&4&-0.006~761~(3)&$1 \times 10^8$ &150   \\
$\Delta M_{2,P6e(P2)}^{(eee)}$&8&0.072~639~(9)&$1 \times 10^8$ &150   \\
$\Delta M_{2,P6f(P2)}^{(eee)}$&4&-0.051~169~(4)&$1 \times 10^8$ &150   \\
$\Delta M_{2,P6g(P2)}^{(eee)}$&2&0.028~956~(4)&$1 \times 10^8$ &150   \\
$\Delta M_{2,P6h(P2)}^{(eee)}$&2&0.003~673~(3)&$1 \times 10^8$ &150   \\
  \\
$\Delta M_{2,P6a(P2)}^{(eem)}$&4& 0.000~99~(7)\times 10^{-4}&$1 \times 10^7$ &50   \\
$\Delta M_{2,P6b(P2)}^{(eem)}$&2& 0.001~08~(5)\times 10^{-4}&$1 \times 10^7$ &50   \\
$\Delta M_{2,P6c(P2)}^{(eem)}$&4&-0.006~60~(5)\times 10^{-4}&$1 \times 10^7$ &50   \\
$\Delta M_{2,P6d(P2)}^{(eem)}$&4&-0.014~12~(14)\times 10^{-4}&$1 \times 10^7$ &50   \\
$\Delta M_{2,P6e(P2)}^{(eem)}$&8& 0.463~55~(84)\times 10^{-4}&$1 \times 10^7$ &50   \\
$\Delta M_{2,P6f(P2)}^{(eem)}$&4&-0.476~36~(57)\times 10^{-4}&$1 \times 10^7$ &50   \\
$\Delta M_{2,P6g(P2)}^{(eem)}$&2& 0.140~65~(50)\times 10^{-4}&$1 \times 10^7$ &50   \\
$\Delta M_{2,P6h(P2)}^{(eem)}$&2&-0.198~15~(63)\times 10^{-4}&$1 \times 10^7$ &50   \\
  \\
$\Delta M_{2,P6a(P2)}^{(eme)}$&4& 0.073~23~(7)\times 10^{-4}&$1 \times 10^7$ &50   \\
$\Delta M_{2,P6b(P2)}^{(eme)}$&2& 0.047~15~(5)\times 10^{-4}&$1 \times 10^7$ &50   \\
$\Delta M_{2,P6c(P2)}^{(eme)}$&4& 0.015~22~(6)\times 10^{-4}&$1 \times 10^7$ &50   \\
$\Delta M_{2,P6d(P2)}^{(eme)}$&4&-0.061~12~(6)\times 10^{-4}&$1 \times 10^7$ &50   \\
$\Delta M_{2,P6e(P2)}^{(eme)}$&8& 0.398~96~(16)\times 10^{-4}&$1 \times 10^7$ &50   \\
$\Delta M_{2,P6f(P2)}^{(eme)}$&4&-0.150~36~(5)\times 10^{-4}&$1 \times 10^7$ &50   \\
$\Delta M_{2,P6g(P2)}^{(eme)}$&2& 0.122~27~(5)\times 10^{-4}&$1 \times 10^7$ &50   \\
$\Delta M_{2,P6h(P2)}^{(eme)}$&2& 0.048~79~(4)\times 10^{-4}&$1 \times 10^7$ &50   \\
\end{resulttable}
\end{ruledtabular}
\end{table}
\renewcommand{\arraystretch}{1}


\subsection{Set ${\rm I}(h)^{(eem)}$}
\label{subsec:setIheem}

The contribution of Set I($h$) to $a_e$,
in which the inner vacuum-polarization loop consists of muon,
is given by
\begin{eqnarray}
a_e^{(10)} [{\rm I}(h)^{(eem)}]&=& \sum_{i=a}^h \Delta M_{2,P6i(P2)}^{(eem)}
\nonumber   \\
 &-&4 \Delta B_2 \Delta M_{2,P4(P2)}^{(eem)} 
  -4 \Delta B_{2,P2}^{(em)} \Delta M_{2,P4}^{(ee)} 
\nonumber   \\
 &+& 10 \Delta B_2 \Delta B_{2,P2}^{(em)} M_{2,P2}^{(ee)}
- 2 (\Delta L_{4,P2}^{(em)} + \Delta B_{4,P2}^{(em)}) M_{2,P2}^{(ee)}.
\label{set1heem}
\end{eqnarray}
Numerical values of
$\Delta M_{2,P6i(P2)}^{(eem)}$ 
are listed in Table III,
and auxiliary quantities
are listed in Tables II.
Inserting these values 
in Eq. (\ref{set1heem}) we obtain 
\begin{equation}
a_e^{(10)} [{\rm I}(h)^{(eem)}] = -0.233~5~(13)\times 10^{-4}.
\label{eq:Iheem}
\end{equation}
%


\subsection{Set ${\rm I}(h)^{(eme)}$}
\label{subsec:setIheme}

The contribution of Set I($h$) to $a_e$,
in which the outer vacuum-polarization consists of muon loop,
is given by
\begin{eqnarray}
a_e^{(10)} [{\rm I}(h)^{(eme)}] &=& \sum_{i=a}^h \Delta M_{2,P6i(P2)}^{(eme)}
\nonumber   \\
 &-&4 \Delta B_2 \Delta M_{2,P4(P2)}^{(eme)} 
  -4 \Delta B_{2,P2}^{(me)} \Delta M_{2,P4}^{(em)} 
\nonumber   \\
 &+& 10 \Delta B_2 \Delta B_{2,P2}^{(me)} M_{2,P2}^{(em)}
- 2 (\Delta L_{4,P2}^{(me)} + \Delta B_{4,P2}^{(me)}) M_{2,P2}^{(em)}.
\label{set1heme}
\end{eqnarray}
%
Numerical values of
$\Delta M_{2,P6i(P2)}^{(eme)}$ 
are listed in Table III,
and auxiliary quantities
are listed in Tables II.
Inserting these values 
in Eq. (\ref{set1heme}) we obtain 
\begin{equation}
a_e^{(10)} [{\rm I}(h)^{(eme)}] = 0.127~9~(2)\times 10^{-4}.
\label{eq:Iheme}
\end{equation}

Terms for $(emm)$, $(eet)$, etc. will be even smaller.
Thus, they are of no physical significance at present.
Of course it is  easy to evaluate them, if needed.


\subsection{Contribution to muon $g\!-\!2$ from Set ${\rm I}(h)$}
\label{subsec:setIhmee}

FORTRAN programs for the electron $g\!-\!2$ can be readily converted
to the muon case by changing one or two parameters.
In the case $(mee)$ the renormalized contribution of Set I($h$) to $a_\mu$ is given by
\begin{eqnarray}
a_\mu^{(10)} [{\rm I}(h)^{(mee)}] &=& \sum_{i=a}^h \Delta M_{2,P6i(P2)}^{(mee)}
\nonumber   \\
 &-&4 \Delta B_2 \Delta M_{2,P4(P2)}^{(mee)} -4 \Delta B_{2,P2}^{(ee)} \Delta M_{2,P4}^{(me)} 
\nonumber   \\
 &+& 10 \Delta B_2 \Delta B_{2,P2}^{(ee)} M_{2,P2}^{(me)}
- 2 (\Delta L_{4,P2}^{(ee)} + \Delta B_{4,P2}^{(ee)}) M_{2,P2}^{(me)}.
\label{set1hmu}
\end{eqnarray}
%
%
%

Substituting numerical values listed in Tables \ref{tablexxaux} and \ref{tableyymee}
in Eq. (\ref{set1hmu}) we obtain 
\begin{equation}
a_\mu^{(10)} [{\rm I}(h)^{(mee)}]= -0.790~83~(59).
\label{eq:amuhee}
\end{equation}

\renewcommand{\arraystretch}{0.80}
\begin{table}
\begin{ruledtabular}
\caption{ Contributions of diagrams
of Set I($h$) of Fig.2 to $a_\mu$ with $(mee)$, $(mem)$, and $(mme)$.
$n_F$ is the number of Feynman diagrams represented by the integral.
All integrals are evaluated in double precision.
  \\
\label{tableyymee}
}
\begin{resulttable}
  \\
$\Delta M_{2,P6a(P2)}^{(mee)}$&4&  4.039~70~(23)&$1 \times 10^9$ &200   \\
$\Delta M_{2,P6b(P2)}^{(mee)}$&2&  2.198~75~(8) &$1 \times 10^9$ &200   \\
$\Delta M_{2,P6c(P2)}^{(mee)}$&4&  1.738~74~(9) &$1 \times 10^9$ &200   \\
$\Delta M_{2,P6d(P2)}^{(mee)}$&4& -3.303~39~(8) &$1 \times 10^9$ &200   \\
$\Delta M_{2,P6e(P2)}^{(mee)}$&8&  7.478~07~(29)&$1 \times 10^9$ &200   \\
$\Delta M_{2,P6f(P2)}^{(mee)}$&4&-12.321~11~(19)&$1 \times 10^9$ &200   \\
$\Delta M_{2,P6g(P2)}^{(mee)}$&2& -1.867~04~(23)&$1 \times 10^9$ &200   \\
$\Delta M_{2,P6h(P2)}^{(mee)}$&2&  6.007~99~(9) &$1 \times 10^9$ &200   \\
  \\ 
 $\Delta M_{2,P6a(P2)}^{(mem)} $  & 4 &     0.003~32~(13)   &  $ 1 \times 10^7 $  & 50    \\
 $\Delta M_{2,P6b(P2)}^{(mem)} $  & 2 &     0.002~236~(77)  &  $ 1 \times 10^7 $  & 50    \\
 $\Delta M_{2,P6c(P2)}^{(mem)} $  & 4 &     0.002~148~(25)  &  $ 1 \times 10^7 $  & 50    \\
 $\Delta M_{2,P6d(P2)}^{(mem)} $  & 4 &    -0.000~639~(80)  &  $ 1 \times 10^7 $  & 50    \\
 $\Delta M_{2,P6e(P2)}^{(mem)} $  & 8 &     0.289~64~(50)   &  $ 1 \times 10^7 $  & 50    \\
 $\Delta M_{2,P6f(P2)}^{(mem)} $  & 4 &    -0.231~04~(18)   &  $ 1 \times 10^7 $  & 50    \\
 $\Delta M_{2,P6g(P2)}^{(mem)} $  & 2 &    -0.047~18~(51)   &  $ 1 \times 10^7 $  & 50    \\
 $\Delta M_{2,P6h(P2)}^{(mem)} $  & 2 &     0.014~71~(7)    &  $ 1 \times 10^7 $  & 50    \\
  \\
 $\Delta M_{2,P6a(P2)}^{(mme)} $  & 4 &     0.206~70~(19)   &  $ 1 \times 10^7 $  & 50    \\
 $\Delta M_{2,P6b(P2)}^{(mme)} $  & 2 &     0.134~50~(15)   &  $ 1 \times 10^7 $  & 50    \\
 $\Delta M_{2,P6c(P2)}^{(mme)} $  & 4 &     0.036~45~(14)   &  $ 1 \times 10^7 $  & 50    \\
 $\Delta M_{2,P6d(P2)}^{(mme)} $  & 4 &    -0.169~33~(13)   &  $ 1 \times 10^7 $  & 50    \\
 $\Delta M_{2,P6e(P2)}^{(mme)} $  & 8 &     1.077~82~(44)   &  $ 1 \times 10^7 $  & 50    \\
 $\Delta M_{2,P6f(P2)}^{(mme)} $  & 4 &    -0.451~81~(13)   &  $ 1 \times 10^7 $  & 50    \\
 $\Delta M_{2,P6g(P2)}^{(mme)} $  & 2 &     0.281~06~(11)   &  $ 1 \times 10^7 $  & 50    \\
 $\Delta M_{2,P6h(P2)}^{(mme)} $  & 2 &     0.146~48~(9)    &  $ 1 \times 10^7 $  & 50    \\
  \\
\end{resulttable}
\end{ruledtabular}
\end{table}
\renewcommand{\arraystretch}{1}

Contributions of terms of type $(mme)$  
and $(mem)$ are
\begin{equation}
a_\mu^{(10)} [{\rm I}(h)^{(mme)}] = 0.253~70~(57),
\label{eq:amuhme}
\end{equation}
and
\begin{equation}
a_\mu^{(10)} [{\rm I}(h)^{(mem)}] = -0.031~48~(62),
\label{eq:amuhem}
\end{equation}
respectively. 
Contributes involving tau lepton is small but not negligible.
Contributions of terms of type $(met)$, $(mte)$, $(mmt)$, $(mtm)$, $(mtt)$ are
\begin{align}
& a_\mu^{(10)}[{\rm I}(h)^{(met)}] = -0.001~303~(17),
\label{eq:amuhet}
\\
& a_\mu^{(10)}[{\rm I}(h)^{(mte)}] =0.003~281~(5),
\label{eq:amuhte}
\\
& a_\mu^{(10)}[{\rm I}(h)^{(mmt)}] = -0.000~612~(4),
\label{eq:amuhmt}
\\
& a_\mu^{(10)}[{\rm I}(h)^{(mtm)}] =0.000~746~(2),
\label{eq:amuhtm}
\\
& a_\mu^{(10)}[{\rm I}(h)^{(mtt)}] = 0.000~026~(1),
\label{eq:amuhtt}
\end{align}

respectively.


\section{Summary and discussion}
\label{sec:summary}

The contribution of Set I($g$) to the electron $g\!-\!2$,
including mass-depending terms where one of lepton loops
is a muon loop, is the sum of 
(\ref{eq:Igeee}),
(\ref{eq:Igeem}), and
(\ref{eq:Igeme}):
\begin{equation}
a_e^{(10)} [{\rm I}(g)] = 0.028~592~(6) \left ( \frac{\alpha}{\pi}\right )^5.
\label{eq:Igall-e}
\end{equation}

The contribution of Set I($h$) to the electron $g\!-\!2$,
including mass-depending terms where one of lepton loops
is a muon loop, is the sum of 
(\ref{eq:Iheee}),
(\ref{eq:Iheme}), and
(\ref{eq:Iheem}):
\begin{equation}
a_e^{(10)} [{\rm I}(h)] = 0.001~685~(13) \left ( \frac{\alpha}{\pi}\right )^5.
\label{eq:Ihall-e}
\end{equation}

Thus the contributions of Set I($g$) and Set I($h$) to the electron $g\!-\!2$
are small and within computational uncertainties
of several other gauge-invariant sets \cite{kn1}.

The contribution of Set I($g$) to the muon $g\!-\!2$,
including terms where lepton loops are  electrons, an electron and  a muon,
an electron and a tau lepton, or muon loops, is the sum of  
(\ref{eq:amugee}),
(\ref{eq:amugme}), 
(\ref{eq:amugem}),
(\ref{eq:amuget}),
(\ref{eq:amugte}), 
(\ref{eq:amugmt}),
(\ref{eq:amugtm}),
(\ref{eq:amugtt}), and
(\ref{eq:Igeee}):
\begin{equation}
a_\mu^{(10)} [{\rm I}(g)] = 2.630~86~(37) \left ( \frac{\alpha}{\pi}\right )^5,
\label{eq:Igall-mu}
\end{equation}
noting that Eq.~(\ref{eq:Igeee}) holds for the case ($mmm$), too.

The contribution of Set I($h$) to the muon $g\!-\!2$,
including terms where lepton loops are  electrons, an electron and  a muon,
an electron and a tau lepton, or muon loops, is the sum of  
(\ref{eq:amuhee}),
(\ref{eq:amuhme}), 
(\ref{eq:amuhem}), 
(\ref{eq:amuhet}),
(\ref{eq:amuhte}), 
(\ref{eq:amuhmt}),
(\ref{eq:amuhtm}),
(\ref{eq:amuhtt}), and
(\ref{eq:Iheee}):
\begin{equation}
a_\mu^{(10)} [{\rm I}(h)] =  - 0.564~8~(11) \left ( \frac{\alpha}{\pi}\right )^5,
\label{eq:Ihall-mu}
\end{equation}
noting that Eq. (\ref{eq:Iheee}) holds for the case ($mmm$), too.

Thus the contributions of  Sets I($g$) and I($h$) to the muon $g\!-\!2$
are also very small compared with the total contribution
663 (20) $(\alpha/\pi)^5$ 
of 18 gauge-invariant sets, which 
include all dominant terms containing light-by-light-scattering
subdiagrams and/or vacuum-polarization subdiagrams \cite{kn2}. 
Concerning the total contribution to the muon $g\!-\!2$, see also \cite{Kataev}.

Recently an analytic value of the asymptotic contribution of
the sum of Set I($g$)
and Set I($h$) to $a_\mu$ has been obtained:\cite{chetyrkin}
\begin{equation}
a_\mu^{(10)} [{\rm I}(g+h)^{(mee)}:\text{analytic-asymptotic}] = 1.501~12 \left (
\frac{\alpha}{\pi} \right )^5.
\label{anal-asym}
\end{equation}
This is in fair agreement with our result
\begin{equation}
a_\mu^{(10)} [{\rm I}(g+h)^{(mee)}] = 1.560~70~(64) \left (
\frac{\alpha}{\pi} \right )^5.
\label{numericalamu}
\end{equation}
Note, however, that the difference 
between (\ref{anal-asym}) and (\ref{numericalamu}) is 93 times 
larger than the estimated uncertainty of numerical integration.
This may be attributed to the ${\cal O} (m_e/m_\mu)$ term
not included in the analytic result.



\begin{acknowledgments}

This work  is supported in part by JSPS Grant-in-Aid 
for Scientific Research (C) No. 19540322.
T. K.'s work is supported by the U. S. National Science Foundation
Grant No. PHY-0355005.
T. K. thanks RIKEN for the hospitality extended to him while 
part of this work was carried out.
M.~H. is also supported in part by JSPS Grant-in-Aid 
for Scientific Research (C) No. 20540261. 
Numerical computations were partly conducted on 
the RIKEN Super Combined Cluster System (RSCC).

\end{acknowledgments}


\end{document}